\documentclass[usenatbib,useAMS]{mnras}

\usepackage{newtxtext,newtxmath}
\usepackage{amsmath}
\usepackage{bm}

\usepackage[T1]{fontenc}
\usepackage{ae,aecompl}

\usepackage{multicol}
\usepackage{graphicx}
\usepackage{pdflscape}
\usepackage{color}
\usepackage{xcolor}

\newcommand{\codefont}[1]{{\texttt{#1}}}
\newcommand{\Mpch}{\mathrm{\,Mpc\,}h^{-1}}

\title[Quantifying the Rarity of the Local Super-Volume]{Quantifying the Rarity of the Local Super-Volume}

\author[S. Stopyra et al.]{Stephen Stopyra$^{1}$%
	\thanks{Contact e-mail: \href{mailto:s.stopyra@ucl.ac.uk}{s.stopyra@ucl.ac.uk}},
	Hiranya V. Peiris$^{1,2}$, Andrew Pontzen$^{1}$, Jens Jasche$^{2}$, Priyamvada Natarajan$^{3,4,5}$
	\\
	$^{1}$Department of Physics and Astronomy, University College London, London WC1E 6BT, UK\\
	$^{2}$The Oskar Klein Centre for Cosmoparticle Physics, Department of Physics,\\ Stockholm University, AlbaNova, Stockholm SE-106 91, Sweden\\
	$^{3}$Department of Astronomy, Yale University, 52 Hillhouse Avenue, New Haven, CT 06511, USA\\
	$^{4}$Department of Physics, Yale University, P.O. Box 208121, New Haven, CT 06520, USA\\
	$^{5}$Black Hole Initiative, 20 Garden Street, Cambridge, MA 02138, USA}

\date{Accepted XXX. Received YYY; in original form ZZZ}
\pubyear{2021}

\begin{document}
\label{firstpage}
\pagerange{\pageref{firstpage}--\pageref{lastpage}}
 \maketitle

	\begin{abstract}
 We investigate the extent to which the number of clusters of mass exceeding $10^{15}\,M_{\odot}\,h^{-1}$ within the local super-volume ($<135\Mpch$) is compatible with the standard $\Lambda$CDM cosmological model. Depending on the mass estimator used, we find that the observed number $N$ of such massive structures can vary between $0$ and $5$. Adopting $N=5$ yields $\Lambda$CDM likelihoods as low as $2.4\times 10^{-3}$ (with $\sigma_8=0.81$) or $3.8\times 10^{-5}$ (with $\sigma_8=0.74$). However, at the other extreme ($N=0$), the likelihood is of order unity. Thus, while potentially very powerful, this method is currently limited by systematic uncertainties in cluster mass estimates. This motivates efforts to reduce these systematics with additional observations and improved modelling. 
	\end{abstract}


	\section{Introduction}
	\label{sec:introduction}

There is a long history of testing the Copernican principle, and the $\Lambda$CDM model more broadly, by searching for structures or regions in the Universe that appear to be unlikely to arise by chance. Previous studies have focused on the abundance of individual extreme structures, including clusters such as the Sloan Great Wall or Shapley supercluster~\citep{nichol2006effect,sheth2011unusual}, and the Local Void~\citep{xie2014local}. The compatibility of individual structures such as these with $\Lambda$CDM can be quantified using extreme value statistics such as the Gumbel distribution~\citep{gumbel1958statistics}. Because the predicted number of halos declines exponentially with mass, even a single example of an unexpectedly high-mass cluster can be a significant challenge to $\Lambda$CDM.  Recent works using these techniques include~\citet{davis2011most} and~\citet{harrison2011exact,harrison2012testing}.

However, the statistical power of individual objects is always limited, especially if their mass is observationally uncertain. A more powerful approach is to consider the likelihood of multiple massive structures coexisting in a small volume. $\Lambda$CDM provides a prediction for the expected number density of clusters above a given mass threshold; combining this with a statistical model of fluctuations away from the mean, one can quantify how likely it is to find the observed number of clusters in a given volume.  The results can in principle be used to place constraints on extensions to $\Lambda$CDM such as primordial non-Gaussianity~\citep{loverde2011non}.  
	
In this work, we consider the number of clusters exceeding the threshold mass $10^{15}\,M_{\odot}\,h^{-1}$ in the local region $<135\Mpch$ (approximately $z\leq 0.046$). We will refer to this volume as the {\sl local super-volume}. To obtain a sensitive test of $\Lambda$CDM, the choices of mass threshold and volume are coupled; for maximal statistical sensitivity, we have adopted values which set an expectation of $O(1)$ above-threshold clusters. We use mass estimates from a variety of methods, allowing us to assess whether the systematic uncertainties are sufficiently well-controlled to obtain a reliable likelihood under the assumption of $\Lambda$CDM.

In Sec.~\ref{sec:estimation} we outline our method for quantifying the rarity of a volume containing multiple massive clusters. Sec.~\ref{sec:masses} describes available mass estimation methods and discusses the available estimates for clusters of interest in the local super-volume. We present our results on the rarity of the local super-volume in Sec.~\ref{sec:results}. In Sec.~\ref{sec:discussion}, we discuss the impact of possible systematics and considerations for improving this method in the future.
	
	\section{Methods}
	\label{sec:estimation}

In this section, we  describe how the halo mass function can be used to place constraints on specific regions, such as the local super-volume. By default, we assume a flat $\Lambda$CDM cosmology with the {\sl Planck} 2018 cosmological parameters~\citep{aghanim2020planck}. This corresponds to a matter density $\Omega_m = 0.315$, a matter power spectrum normalisation $\sigma_8 = 0.811$, and $h = 0.674$ for the Hubble rate, $H_0 = 100 h \mathrm{\,kms}^{-1}\mathrm{Mpc}^{-1}$. We will also explore the effect of lowering the power spectrum normalisation to agree with weak lensing results, adopting $\sigma_8=0.741$ \citep{asgari2021kids} while fixing $\Omega_m$ and $h$ to the {\sl Planck} values.
	
The expected number of clusters, $N_{\mathrm{exp}}$, within volume $V$ and with mass $M \geq M_{\mathrm{thresh}}$ is obtained by integrating the halo mass function, $\mathrm{d}n(M)/\mathrm{d}M$,
	\begin{equation}
	N_{\mathrm{exp}} = V\int_{M_{\mathrm{thresh}}}^{\infty}\frac{\mathrm{d}n(M)}{\mathrm{d}M}\mathrm{d}M.\label{eq:Nexp}
	\end{equation}
To quantify the likelihood of the number of  clusters actually observed in a given volume, we additionally require a statistical model for fluctuations away from this expectation value.
	
\begin{figure}
	\centering
	\includegraphics{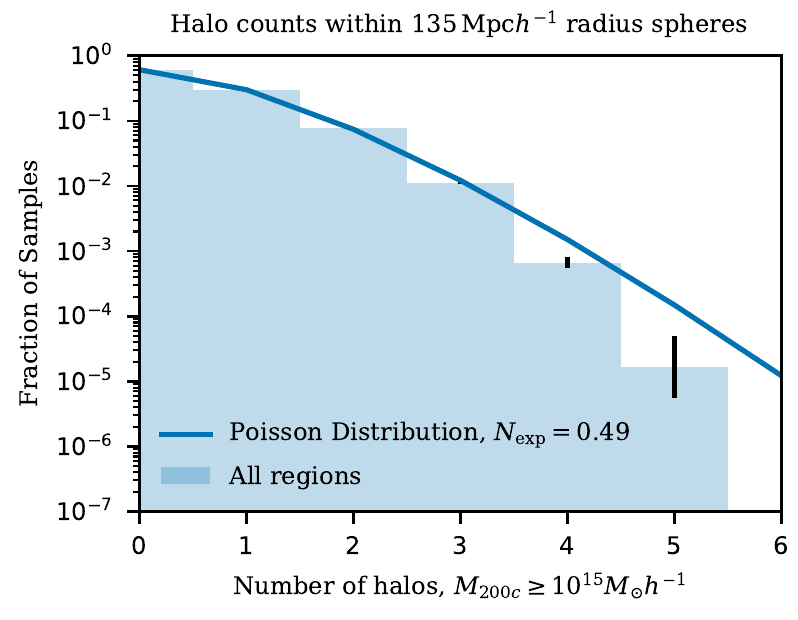}
	\caption{\label{fig:poisson}
Fraction of randomly-selected $135\Mpch$ spheres in six simulations with a given number of halos above $M_{200c} \geq 10^{15}M_{\odot}h^{-1}$, compared with a Poisson distribution with mean cluster count fixed to be the same as that of the simulations.}
\end{figure}

Specifically, we assume that the likelihood of observing $N$ clusters follows a Poisson distribution with mean $N_{\mathrm{exp}}$, i.e.
\begin{equation}
	\mathcal{L}(N|N_{\mathrm{exp}}) = \frac{N_{\mathrm{exp}}^N \, \mathrm{e}^{-N_{\mathrm{exp}}}}{N!}.\label{eq:sf}
	\end{equation}
To test the validity of this assumption, we performed six $512^3$-particle $\Lambda$CDM simulations with a side-length of $677.7\Mpch$, from which we randomly extracted spheres of the same size as the local super-volume. We confirmed that the distribution of the number of halos with masses above $10^{15}M_{\odot}h^{-1}$ was well approximated by a Poisson distribution, as shown in Fig.~\ref{fig:poisson}. The simulated distribution shows marginally lower probabilities in the high-$N$ tail, meaning our Poisson likelihoods should be regarded as an upper limit.

\begin{figure*}
	\centering
	\includegraphics[width=\textwidth]{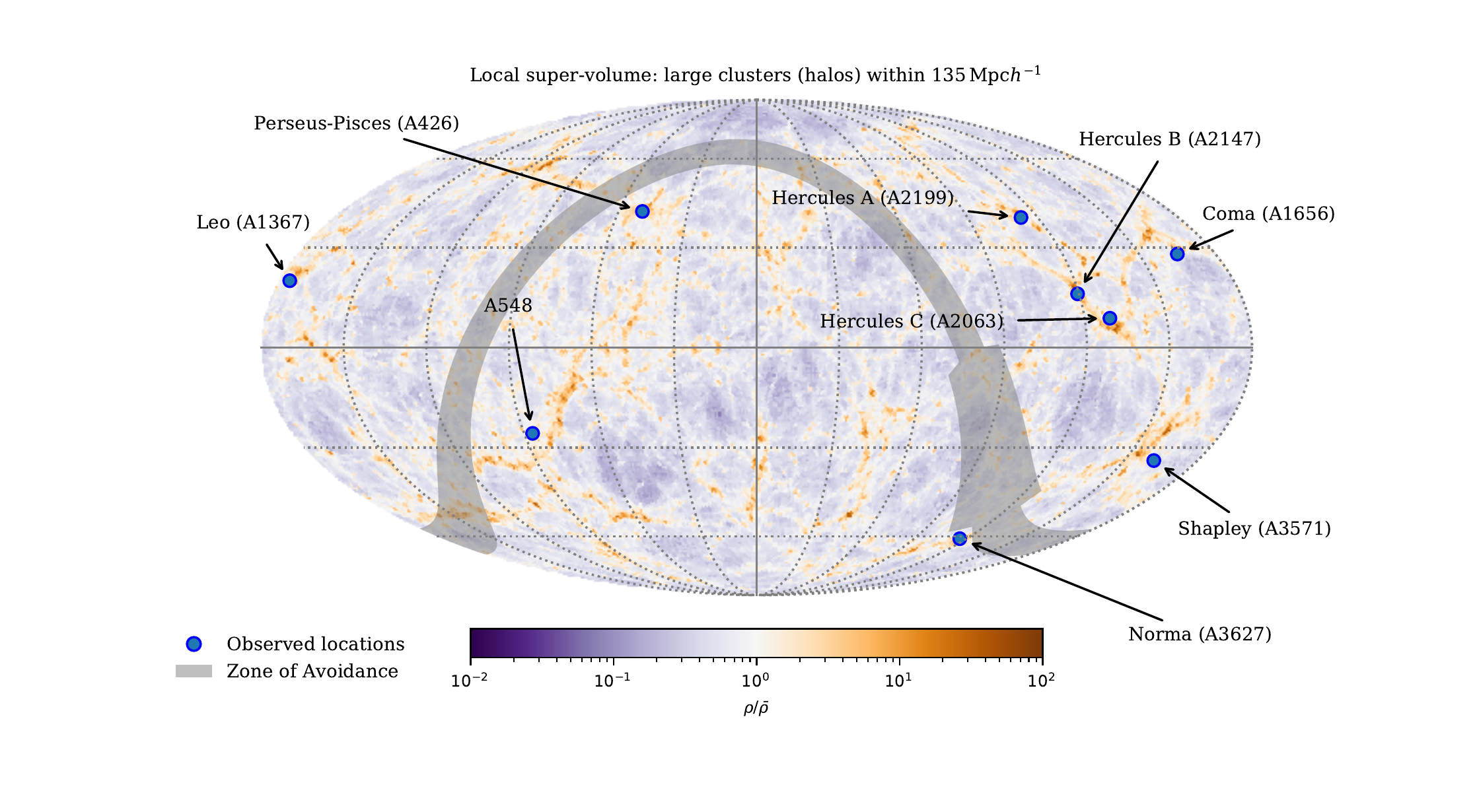}
	\caption{\label{fig:skyHalos}Mollweide projection of the full sky in equatorial co-ordinates, showing the locations of the nine clusters considered in this work (blue circles). The projected density out to $135\Mpch$ is also shown, as inferred by the \codefont{BORG} algorithm~\citep{jasche2019physical} using the 2M++ galaxy catalogue~\citep{2mppPaper}. The zone of avoidance is shown in grey and defined in Galactic co-ordinates as $-5^{\circ} < l < 5^{\circ}$, except within the region of the Galactic centre, $-30^{\circ} < b < 30^{\circ}$, where it includes $-10^{\circ} < l < 10^{\circ}$.}
\end{figure*}

This method was then used to quantify the rarity of the local super-volume by counting the number of clusters with masses above $>10^{15}\,M_{\odot}\,h^{-1}$.  Because the likelihood function is highly sensitive to $N$ (decreasing rapidly when $N > N_{\mathrm{exp}}$), it is essential to obtain accurate estimates of the cluster masses; we turn to this crucial issue in the next section. 

Another important consideration is the sensitivity of $N_{\mathrm{exp}}$ to the choice of halo mass function. To obtain accurate estimates of the abundance of high-mass clusters, it is necessary to properly account for the effects of large-scale modes~\citep{park2012challenge}. In particular, \citet{2015JKAS...48..213K} compared several mass functions in the literature to the halo number counts in the Horizon Run 4 simulation, which has a very large box size ($\sim3\mathrm{\,Gpc}$). They found that most mass functions inaccurately predict the number of high-mass halos, which can significantly affect $N_{\mathrm{exp}}$ and hence the likelihood in Eq.~(\ref{eq:sf}). 

Using mass functions calibrated with large-volume simulations is therefore essential to provide accurate likelihood estimates. For the purposes of this work we use a mass function calibrated using the Horizon Run 4 simulations~\citep{2015JKAS...48..213K}; we convert between the Friends-of-Friends masses used by the Horizon Run 4 mass function, and the spherical-overdensity masses used in this work, by using the relation of~\citet{more2011overdensity}, with concentrations given by the~\citet{bhattacharya2013dark} concentration-mass relationship.
	
As previously outlined, the choice of mass threshold, $M_{\mathrm{thresh}}$, is also important in interpreting the final likelihood. In this work, we consider the number of clusters with $M_{200c} \geq 10^{15}M_{\odot}h^{-1}$, where $M_{200c}$ is the mass within a radius such that the average density is 200 times the critical density of the Universe. This threshold is somewhat arbitrary, but is chosen for two reasons: it corresponds to an expected abundance of $O(1)$ in a volume the size of the local super-volume; further, few clusters are found above this mass threshold in the Universe at large~\citep{ade2016planck,hilton2018atacama}, since it is around the scale of the largest structures that have had time to viralise by redshift $z = 0$~\citep{1974ApJ...187..425P}. Consequently, the halo mass function above this mass is poorly constrained observationally. A significantly lower mass threshold (for example, $5\times 10^{14}M_{\odot}h^{-1}$) would have $O(10)$ or more clusters in the local super-volume, making the computed likelihood insensitive to the addition of one or two extreme-mass objects, while a significantly higher threshold would run into essentially the same limitations in statistical power that arise when studying individual objects.

	\section{Cluster Mass Estimates}
	\label{sec:masses}

	We now turn to obtaining estimates for the masses of the most extreme local clusters. We briefly review different mass estimation methods -- with a view to highlighting advantages and current limits -- and discuss estimates available in the literature for massive local clusters and super-clusters.
	
	The clusters on which we focus are shown in in Fig.~\ref{fig:skyHalos}, along with their Abell catalogue numbers~\citep{abell1989catalog}. These clusters are consistently represented as massive halos in reconstructions of the local super-volume that make use of Bayesian-Origin-Reconstruction from Galaxies (\codefont{BORG}). Specifically, the clusters we have selected correspond to the nine most massive local structures in a reconstruction performed by~\citet{jasche2019physical} of the local super-volume, using the 2M++ galaxy catalogue~\citep{2mppPaper} at high signal-to-noise ratio out to $135\Mpch$.  In the future, using improved forward-modelling, \codefont{BORG} itself could be used to give independent mass estimates for these clusters; however in this work we only use more traditional mass estimates.
	
Mass estimates for these clusters taken from the literature are collated in Fig.~\ref{fig:observations}. All mass estimates have been converted to $M_{200c}$ masses using the concentration-mass relationship of~\citet{bhattacharya2013dark}. Since much of the literature uses $M_{500c}$ masses, the typical correction is an increase in mass by $\sim 30$\%, with a maximum correction of 31\%. Assuming that all halos follow the mean relationship will introduce some error into these extrapolations (included in the error bars on Fig.~\ref{fig:observations}), but cannot account for the large discrepancies between different mass estimators that we highlight below.
	
	\subsection{Review of Mass Estimation Methods}
	\label{sec:methods}

	The most common methods used to estimate cluster masses fall into four main categories: dynamical estimates using the virial theorem~\citep{merritt1987distribution}; weak lensing~\citep{1994ApJ...427L..83B,1994ApJ...437...56F}; X-ray masses~\citep{1996ApJ...469..494E}; and the Sunyaev Zel'dovich (SZ) method~\citep{sunyaev1970spectrum,sunyaev1980microwave}. In this section we briefly review each method in turn.
	
	Based on the virial theorem, \citet{girardi1998optical} showed the virial mass, $M_V$, may be estimated using
	\begin{equation}
	M_V = \frac{\langle v^2\rangle}{\langle r^{-1}F(r)\rangle},
	\end{equation}
	where $\langle v^2\rangle$ is the average velocity dispersion and $F(r)$ is the fraction of mass of the cluster that lies within the radius $r$~\citep{merritt1987distribution}. The fundamental challenge in the virial theorem approach -- and other dynamical methods -- is to approximate the underlying matter distribution, described by $F(r)$. Early studies~\citep[e.g.][]{girardi1998optical} assumed that the dark matter was traced by the galaxy distribution. Later methods~\citep[e.g.][]{lokas2003dark} instead fit the moments of the observed velocity distribution to a Navarro-Frenk-White (NFW) profile~\citep{Navarro:1995iw}. This improves the accuracy of masses since one does not have to assume that the dark matter traces the cluster galaxies. However, it is constrained by the assumption that local
	clusters are well-fit by a spherical NFW profile. Moreover, all dynamical estimates can be inaccurate if the cluster is not in equilibrium, e.g., due to a recent merger~\citep{takizawa2010mass}. To partially mitigate these systematics, one can fit the observed velocity dispersion to a dispersion-mass relationship calibrated on simulations~\citep{2013MNRAS.430.2638M,aguerri2020deep}.
	
	Weak lensing is a commonly-used mass estimation method at high redshift; however it has also  been used locally, with several studies of the Coma cluster~\citep{kubo2007mass,okabe2014subaru,gavazzi2009weak}, A2199~\citep{kubo2009sloan}, and A2063~\citep{sereno2017psz2lens}. A significant source of systematic errors in weak lensing estimates arises from the contamination of the lensing signal from unassociated structures that happen to lie along the line-of-sight. Without sufficient redshift precision, it can be difficult to distinguish cluster members from lensed background galaxies and foreground galaxies, leading to systematic overestimates of mass. 
	
\begin{figure*}
		\centering
		\includegraphics[width=\textwidth]{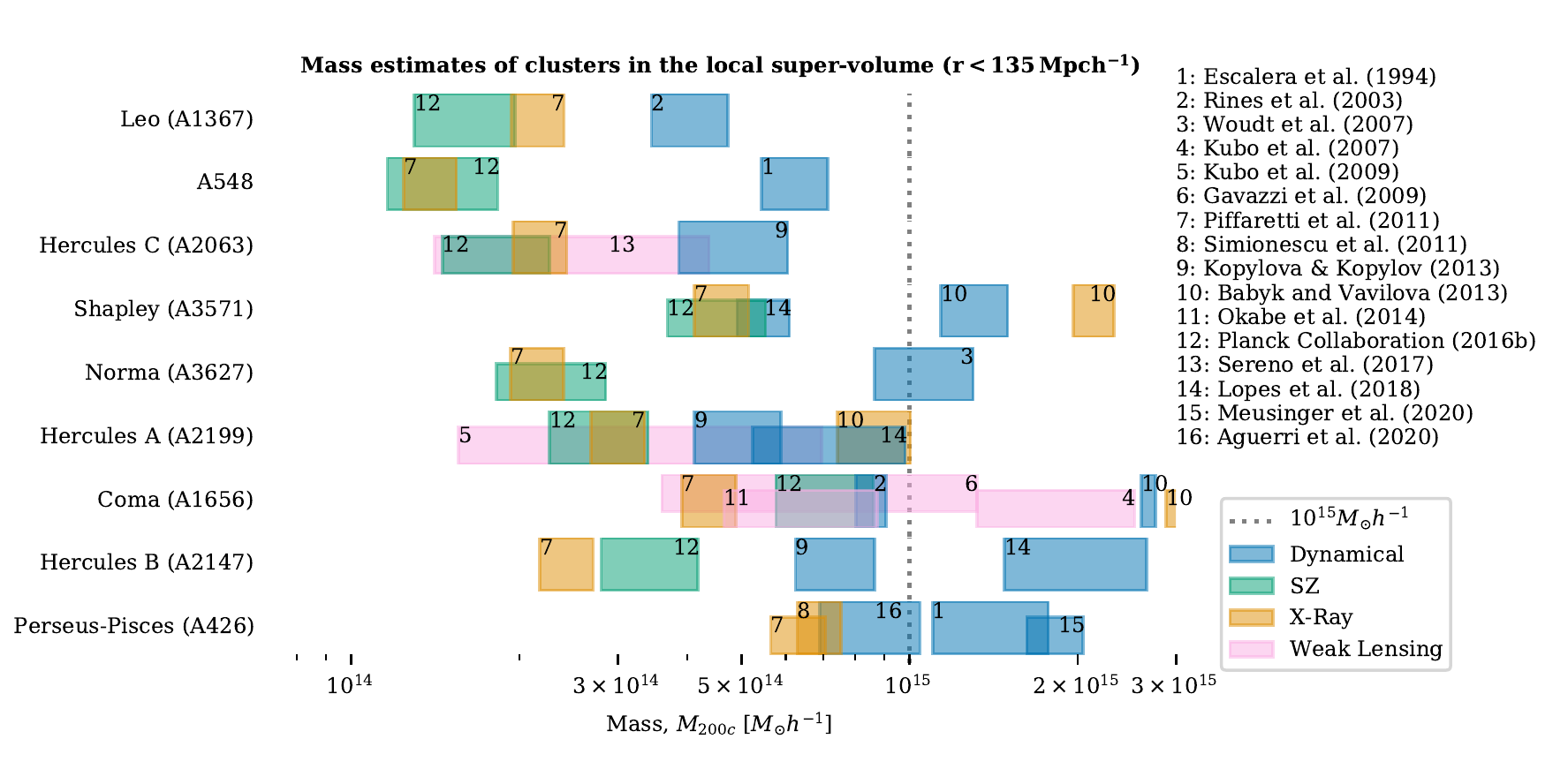}
		\caption{\label{fig:observations}Mass estimates for nine massive clusters in the local super-volume. Shaded regions show the $1\sigma$ bounds on estimates of the cluster mass, including both statistical and any systematic errors that have been accounted for. Numbers indicate the reference for each estimate, while colours indicate the method used in the mass estimate.}
	\end{figure*}

	The X-ray approach~\citep{1996ApJ...469..494E} makes use of thermal bremsstrahlung emitted by  hot cluster gas, generally assuming isothermal hydrostatic equilibrium~\citep{2012RAA....12..973O}. These assumptions are violated in merging clusters, which can lead to significant biases. Feedback effects from the accreting active galactic nuclei (AGN) harboured by the brightest cluster galaxies (BCGs) are expected to redistribute and modulate the mass distribution in the innermost regions of clusters. As evident from X-ray data of the Perseus cluster~\citep{fabian2000chandra}, the choice of integrating the mass within the significantly larger $R_{\mathrm{200c}}$ (the radius such that the average density is 200 times the critical density) mitigates this effect. Overall, there is evidence that the effects of cluster mergers are less significant for X-ray measurements than for dynamical estimates~\citep{takizawa2010mass}. For the results in Fig.~\ref{fig:observations}, we use X-ray masses mainly from the MCXC catalogue~\citep{Piffaretti_2011}, with some estimates from~\citet{babyk2013comparison} and~\citet{simionescu2011baryons}.

	The thermal SZ effect allows mass measurements using the up-scattering of cosmic microwave background (CMB) photons by the hot intracluster medium~\citep{sunyaev1970spectrum,sunyaev1980microwave}. This produces a spectral distortion which can be detected with high-precision measurements of the CMB~\citep{Ade:2015gva}. The SZ masses are determined using a scaling relationship between the Compton parameter, $Y_{SZ}$, and the cluster mass, $M_{SZ}$. This scaling relationship is calibrated using X-ray estimates, $M_X$. The dominant source of uncertainty is the assumed mass-bias, $M_{X} = (1-b)M_{SZ}$, which accounts for biases in X-ray masses such as departure from hydrostatic equilibrium. We use the {\sl Planck} 2015 SZ masses~\citep{Ade:2015gva}; this work estimates $0.7<(1-b)<1.0$. We include the corresponding uncertainty in the error bars shown in Fig.~\ref{fig:observations}.

	\subsection{Individual clusters}
	\label{sec:selection}
	
	In this work, we focus our attention on nine of the most massive clusters in the local super-volume, whose positions are shown in Fig.~\ref{fig:skyHalos}, covering both hemispheres of the sky. We now briefly review what is known about each structure. The mass estimates that we discuss are compiled in Fig.~\ref{fig:observations}.\\
	
\noindent {\bf Perseus-Pisces (A426):} The Perseus-Pisces supercluster is dominated by the rich Abell cluster A426 (also known as the Perseus Cluster). It was one of the first identified superclusters~\citep{1978IAUS...79..241J} and is among the most massive in the local super-volume~\citep{escalera1994structures}. However, there is considerable disagreement in the literature on its mass, with X-ray results \citep{simionescu2011baryons} pointing to a somewhat smaller mass than dynamical estimates \citep{aguerri2020deep}. The latter use a velocity-dispersion-to-mass relationship studied in~\citet{2013MNRAS.430.2638M}, which uses simulations to account for baryonic effects.~\citet{meusinger2020galaxy} find an even larger mass using the virial method, which agrees with earlier virial theorem estimates such as those by~\citet{escalera1994structures}.

\vspace{1em}
	
	\noindent {\bf Hercules A (A2199 \& A2197):}	
	The pair of clusters A2197 and A2199 form the Hercules A portion of the Hercules supercluster system, and is believed to be in the process of merging, according to the dynamical analysis by~\citet{krempec2002binary}. Estimates vary for the mass of the largest member of the pair, A2199: \citet{lopes2018optical} use the virial theorem with a pressure-term correction~\citep{girardi1998optical} to estimate the mass of A2199, obtaining results slightly higher than~\citet{kopylova2013investigation}, but still consistent with them. Both give higher masses than the X-ray estimates of~\citet{Piffaretti_2011} and SZ results from~\citet{Ade:2015gva}. A weak lensing estimate is also available from~\citet{kubo2009sloan}; however this gives a very broad range of possible masses, insufficient to distinguish between the X-ray/SZ and dynamical estimates. An additional X-ray estimate has been provided by~\citet{babyk2013comparison}, who obtain a higher mass than~\citet{Piffaretti_2011}.

		\begin{table*}
		\centering
		\begin{tabular}{c|c|c}
			Cluster Count, $N$, $M_{200c} \geq 10^{15}M_{\odot}h^{-1}$  & Likelihood $(N_{\mathrm{exp}} = 0.94, \sigma_8 = 0.81)$ & Likelihood $(N_{\mathrm{exp}} = 0.37, \sigma_8 = 0.74)$ \\
			\hline
			$0$ & $0.39$ & $0.69$ \\
			$1$ & $0.37$ & $0.25$\\
			$2$ & $0.17$ & $4.6\times 10^{-2}$\\
			$3$ & $5.4\times 10^{-2}$ & $5.7\times 10^{-3}$\\
			$4$ & $1.3\times 10^{-2}$ & $5.2\times 10^{-4}$\\
			$5$ & $2.4\times 10^{-3}$ & $3.8\times 10^{-5}$\\
		\end{tabular}
		\caption{\label{tab:nExp} Likelihood that a randomly-selected region of the $\Lambda$CDM Universe has $N$ clusters of mass $M_{200c} \geq 10^{15}M_{\odot}h^{-1}$. This follows from Eq.~\ref{eq:sf}. $N_{\mathrm{exp}}$ is computed using the Horizon Run 4 mass function~\citep{2015JKAS...48..213K} with the {\sl Planck} 2018 cosmology~\citep{aghanim2020planck} $\Omega_m = 0.315,h = 0.674$. In the first column we show the likelihoods with the {\sl Planck} 2018 value of $\sigma_8 = 0.811$, while in the second column we show the effect of a lower value from KiDS~\citep{asgari2021kids}, $\sigma_8 = 0.741$, for the same $\Omega_m$ and $h$.}
	\end{table*}
	
	\noindent {\bf Hercules B (A2147, A2151 \& A2152):}	
	The group of clusters around A2147 is sometimes known as the Hercules B system, and includes A2151 and A2152. Like the Hercules A system, it is believed to be gravitationally bound and in the process of collapsing~\citep{krempec2002binary,kopylova2013investigation}. We will focus our attention on the largest of these three clusters, A2147: both~\citet{lopes2018optical} and~\citet{kopylova2013investigation} give virial estimates for the mass of this cluster, but disagree on the mass by a factor of $\sim2$. These dynamical estimates are also much higher than X-ray~\citep{Piffaretti_2011} and SZ~\citep{Ade:2015gva} estimates. Given that A2147 interacts with the two nearby clusters A2151 and A2152, the assumption of dynamical equilibrium relied upon by all these methods may be questionable, and require more detailed analysis of the entire system.
	
	\noindent {\bf Hercules C (A2063 \& A2052):}	
	The Hercules super-cluster contains a third major concentration of galaxies centred around the clusters A2063 and A2052, which we dub the Hercules C system to distinguish it from the other groups of Hercules clusters. Like Hercules A and B, this group of clusters appears to be a merging system made up of closely interacting clusters, with slightly lower masses than Hercules A and B. Here, we focus on estimates of the mass of A2063, which is consistently found to be the higher-mass of the two clusters. Our main dynamical results are from~\citet{kopylova2013investigation}, who find a higher mass than the X-ray/SZ results~\citep{Piffaretti_2011,Ade:2015gva}. \citet{sereno2017psz2lens} considered weak lensing of {\sl Planck} SZ clusters, finding results which are compatible with both dynamical and X-ray/SZ estimates. As with the Hercules A and B systems, the effect of the close-interaction with nearby clusters on these mass estimates (in this case A2052) is not well-understood.
	
	\noindent {\bf Coma (A1656):}	
	The Coma super-cluster has two main clusters, A1656 and A1367. The most massive of these, A1656, is known as the Coma cluster and has been widely studied. There have been several attempts to estimate the mass of the Coma cluster using weak lensing:~\citet{kubo2007mass} find a relatively high-mass compared to more recent results~\citep{gavazzi2009weak,okabe2014subaru};~\citet{okabe2014subaru} suggest that this may be because~\citet{kubo2007mass} do not properly account for the lensing effect of unassociated background large scale structure. The more recent weak lensing results are consistent with X-ray/SZ estimates~\citep{Piffaretti_2011,Ade:2015gva}, and also the dynamical results of~\citet{rines2003cairns}.~\citet{babyk2013comparison} compared virial and X-ray estimates of the masses of multiple clusters, including $A1656$. Their X-ray and virial estimates are both much higher than the X-ray/SZ estimates of~\citet{Piffaretti_2011,Ade:2015gva}.
	
	\noindent {\bf Leo (A1367):}	
This is arguably the least massive of the nine clusters we consider. The cluster is believed to have undergone a recent merger~\citep{sun2002chandra}, and so virial and X-ray mass estimates may be inaccurate. In Fig.~\ref{fig:observations} we use the corrected virial mass estimate from~\citet{rines2003cairns}, which produces a much higher mass than the X-ray/SZ estimates~\citep{Piffaretti_2011,Ade:2015gva}.
		
	\noindent {\bf Norma (A3627):}	
	This lies close to the zone of avoidance and appears to be associated with the `Great Attractor'. Due to its location, the cluster is not as well studied as some of the others considered here. X-ray/SZ estimates~\citep{Piffaretti_2011,Ade:2015gva} are in agreement, while a virial mass estimate was made by~\citet{woudt2007norma}, and gives a significantly higher mass. 
	
	\noindent {\bf Shapley (A3571):}	
	This cluster lies in the Shapley concentration. The main Shapley group of clusters is also known to be massive, but most of the rest of its members do not fall within the local super-volume. The mass estimates for A3571 are bimodal. Three results agree despite using different techniques: dynamical ~\citep{lopes2018optical}, X-ray~\citep{Piffaretti_2011}, and SZ~\citep{Ade:2015gva}. However, \citet{babyk2013comparison} use two methods (dynamical and X-ray) to arrive at much higher estimates. We note that, as with A2199 and A1656,~\citet{babyk2013comparison} give significantly higher X-ray mass estimates than~\citet{Piffaretti_2011}. 
	
	\noindent {\bf A548:}	
	This is a cluster that is believed to have significant substructure, centred around two main concentrations~\citep{andreuzzi1998redshift}, and is thus likely to be in the process of undergoing a merger. The mass of the combined system was estimated by~\citet{escalera1994structures} using the virial theorem. X-ray~\citep{Piffaretti_2011} and SZ~\citep{Ade:2015gva} estimates give much lower mass than this dynamical estimate~\citep{escalera1994structures}.
		
	\section{Results}
	\label{sec:results}

We now summarize the implications of these mass estimates for cosmology. As can be seen from Fig.~\ref{fig:observations}, there are considerable variations between different mass estimates for the nine massive clusters we have considered. The error bands show the reported combined statistical and systematic errors for each of the cluster mass estimates, which are in many cases significantly smaller than the variations between estimates. Dynamical estimates favour systematically higher masses than X-ray and SZ estimates, typically by a factor of three or more. In some cases the discrepancy in mass estimates is close to an order of magnitude. 
	
In Table~\ref{tab:nExp}, we show the likelihood of observing different numbers of clusters with $M_{200c}>10^{15}M_{\odot}h^{-1}$ in the local super-volume. This allows us to quantify the implications of discrepant mass measurements for cosmology. We compute the likelihood for two different values of $\sigma_8$: the higher value associated with the {\sl Planck} 2018 cosmology~\citep{aghanim2020planck} and a lower value favoured by the Dark Energy Survey (DES)~\citep{2021arXiv210513549D} and the Kilo-Degree Survey (KiDS)~\citep{asgari2021kids} weak lensing results. If one assumes that SZ masses are reliable, there are no observed clusters above the mass threshold within the local super-volume. If we instead assume the midpoint of the dynamical masses, there are four. In the most extreme interpretation of the collated measurements, one could argue there are five.

In a conservative interpretation, therefore, the local super-volume is completely unremarkable for either choice of $\sigma_8$: the likelihoods are order unity. At the other extreme the $\Lambda$CDM model seems very unlikely, yielding likelihoods as low as $2.4\times 10^{-3}$ for $\sigma_8 = 0.81$. The situation is exacerbated if $\sigma_8$ takes a lower value ($0.74$), yielding a likelihood of $3.8\times 10^{-5}$ at the extreme end. An extension to $\Lambda$CDM that predicts higher $N_{\mathrm{exp}}$ would be strongly favoured, demonstrating the potential power of this test. The test could also provide an additional discriminator for the emerging $\sigma_8$ tension~\citep{asgari2021kids,2021arXiv210513549D}. These possibilities motivate observational and modelling programmes to better understand the physical properties of these nearby massive clusters.

	\section{Discussion}
	\label{sec:discussion}
	
We have performed a literature search to collate as many mass estimates as possible for nearby massive clusters, and illustrated the potential for powerful cosmological tests based upon these results. The current barrier to drawing cosmological conclusions is that the mass estimates are in disagreement. 

The mass bias is one of the dominant systematics underlying $SZ$ and X-ray mass estimates.~\citet{medezinski2018planck} compared {\sl Planck} SZ estimates to weak-lensing results for the same clusters as an independent check on the value of the mass bias, finding consistency with~\citet{Ade:2015gva}. However, mass calibration and cross-checks were all undertaken at high redshift.~\citet{andreon2014important} investigated whether there is evidence for redshift evolution in this bias, finding a modest effect (increasing the mass of some low-redshift clusters by up to 10-15\%). While they did not examine the lowest redshift clusters, again the level of correction is small compared to the variation seen in Fig.~\ref{fig:observations}. While the X-ray and SZ estimates therefore appear to provide a picture of self-consistency, one must bear in mind that these methods have in common strong dynamical and symmetry assumptions, which may not be valid for any particular cluster, even if unbiased for the high-redshift population.  

There are a range of systematic concerns regarding dynamical estimates, which can potentially be addressed using better modelling, such as accounting for cluster sub-structure. A completely different dynamical approach, using large scale structures to infer the cluster masses, is offered by \codefont{BORG}~\citep{jasche2019physical}. Improvements to the forward modelling in the \codefont{BORG} algorithm are required in order to robustly resolve cluster scales; we will pursue this in future work.

There is a dearth of weak lensing studies of these clusters, likely because the necessary accuracy for distances is hard to achieve for nearby structures (our study is restricted to $z\leq 0.046$). Given that the precision of photometric redshifts is frequently at the $\Delta z\sim 0.05$ level, contamination of the lensing signal with non-background and non-cluster member galaxies is a significant problem for low redshift clusters.

In principle, weak lensing is the most robust mass estimator, provided that accurate redshifts to a large number of local galaxies can be obtained. This might be achieved, for example, using a dedicated spectroscopic lensing survey, or the upcoming Local Volume Complete Cluster Survey (LoVoCCS)~\citep{fu2021preliminary}. Meanwhile there exists an intriguing situation where we cannot be sure whether the local super-volume is compatible with the standard cosmological model.

	\section*{Acknowledgements}
	The authors gratefully acknowledge useful discussions with Peter Coles and George Efstathiou regarding systematics in mass estimation methods. The name `local super-volume' was suggested to the authors by Daniel Mortlock and is to our knowledge an original term. SS thanks Corentin Cadiou and Jonathan Davies for helpful discussions. This project has received funding from the European Union's Horizon 2020 research and innovation programme under grant agreement No. 818085 GMGalaxies. SS and AP were supported by the Royal Society. HVP was partially supported by the research project grant `Fundamental Physics from Cosmological Surveys' funded by the Swedish Research Council (VR) under Dnr 2017-04212, and HVP and JJ were partially supported by the research project grant `Understanding the Dynamic Universe' funded by the Knut and Alice Wallenberg Foundation under Dnr KAW 2018.0067. JJ acknowledges support by the Swedish Research Council (VR) under the project 2020-05143 -- `Deciphering the Dynamics of Cosmic Structure'. This work used computing equipment funded by the Research Capital Investment Fund (RCIF) provided by UKRI, and partially funded by the UCL Cosmoparticle Initiative. The authors acknowledge the use of the UCL Kathleen High Performance Computing Facility (Kathleen@UCL), and associated support services, in the completion of this work. This research has made use of the SZ-Cluster Database operated by the Integrated Data and Operation Center (IDOC) at the Institut d'Astrophysique Spatiale (IAS) under contract with CNES and CNRS. 
	
	\section*{Author Contributions}
	
	The main roles of the authors were, based on the CRediT (Contribution Roles Taxonomy) system (\url{https://authorservices.wiley.com/author-resources/Journal-Authors/open-access/credit.html}):
	
	{\bf SS:} data curation; investigation; formal analysis; software; visualisation; writing -- original draft preparation.
	
	{\bf HVP:} conceptualisation; methodology; validation and interpretation; writing -- review and editing.
	
	{\bf AP:} conceptualisation; validation and interpretation; writing -- review and editing; funding acquisition.
	
	{\bf JJ:} data curation; resources; writing -- review.
	
	{\bf PN:} interpretation; writing -- review and editing.

	\section*{Data availability}
	The data underlying this article will be shared on reasonable request to the corresponding author.
	
	\interlinepenalty=10000

	\bibliographystyle{mnras}
	\bibliography{clusters_paper}
\label{lastpage}
	
\end{document}